\documentclass[aps,showpacs,twocolumn,floatfix]{revtex4} 
\usepackage{graphicx}

\begin{document}
   
 

\title{Stabilization of a  (3+1)D  soliton in a Kerr medium  by 
a rapidly oscillating dispersion coefficient}

\author{Sadhan K. Adhikari}
\affiliation{Instituto de F\'{\i}sica Te\'orica, Universidade Estadual
Paulista, 01.405-900 S\~ao Paulo, S\~ao Paulo, Brazil}

\date{\today}

\begin{abstract}

Using the numerical solution of the nonlinear Schr\"odinger equation and a
variational method it is shown that (3+1)-dimensional spatiotemporal
optical solitons  can be stabilized by a rapidly
oscillating dispersion coefficient in a Kerr medium with cubic
nonlinearity. This has immediate consequence in generating
dispersion-managed robust optical soliton in communication as well as
possible stabilized Bose-Einstein condensates in periodic optical-lattice
potential via an effective-mass formulation. We also critically compare
the present stabilization with that obtained by a rapid sinusoidal
oscillation  of the
Kerr nonlinearity parameter.

\end{abstract}
\pacs{42.65.Jx, 42.65.Tg, 05.45.Yv, 03.75.Lm}
\maketitle

\section{Introduction}

After the  prediction of self-trapping
\cite{st} of an
optical beam in a nonlinear medium resulting in an optical soliton
\cite{0,1}, there
have been many
theoretical and experimental studies  to stabilize such
a soliton under different conditions of nonlinearity and dispersion.   
A bright soliton
in (1+1) dimension (D)  in Kerr medium (cubic nonlinearity) is
unconditionally stable for
positive or
self-focusing (SF)
nonlinearity in the nonlinear Schr\"odinger
 equation (NLS) \cite{1}.  However, in (2+1)D and (3+1)D, in homogeneous
bulk Kerr medium one cannot have an unconditionally stable soliton-like
beam \cite{1}. (If the nonlinearity is negative or self-defocusing (SDF),
any initially created soliton always spreads out \cite{1}.)  If the
nonlinearity is positive or of SF type, any initially created highly
localized soliton is unstable  in both (2+1)D and
(3+1)D \cite{1}.  Such a confined wave packet in (3+1)D is often called a
light bullet and represents the extension of a self-trapped optical beam
into the temporal domain \cite{1}. The generation of a light bullet is of
vital importance in soliton-based communication systems. On the
experimental front strong stabilization of (2+1)D discrete vortex solitons
in a SF nonlinear medium 
have been observed
in optically induced
photonic lattices \cite{yyy}. 
Such solitons, however, can only be modeled
by a more complicated nonlinearity in the NLS in contrast to the simple
cubic Kerr nonlinearity considered in this paper.

Recently, through a numerical simulation as well as a variational
calculation based on the NLS it has been shown that a soliton in (2+1)D
\cite{ber,mal}  can be stabilized in a layered
medium if a variable Kerr nonlinearity coefficient is used in different
layers \cite{mal,ber}. A weak modulation of the nonlinearity coefficient
along the propagation direction leads to a reasonable stabilization in
(2+1)D \cite{ber}. A much better stabilization results if the Kerr
coefficient is a layered medium is allowed to vary rapidly between
successive SDF and SF type nonlinearities, i.e., between positive and
negative values \cite{mal}.

However, in recent years  optical soliton formed by a controlled
variation of
dispersion coefficient (dispersion-managed optical
soliton) in a Kerr medium
has been of general
interest in communication \cite{A}. 
A dispersion-managed soliton allows
robust
propagation of pulses and is favored over normal solitons. The recent
stabilized 
optical beam in a nonlinear
waveguide array in (1+1)D \cite{T}, called diffraction managed soliton,
is  quite
similar to the present  dispersion-managed optical
soliton. That problem was described \cite{T} by the one
dimensional NLS  with a periodic dispersion
coefficient. Such a periodic variation of the dispersion
coefficient leads to a greater stability of the soliton during
propagation in (1+1)D \cite{T}. More recently, Abdullaev {\it et al}
\cite{abdulla} have shown by a
variational and numerical solution of the  NLS
that by employing a rapidly varying dispersion coefficient it is
possible to stabilize an optical soliton in (2+1)D over large propagation
distances.

The possibility of the  stabilization of an optical soliton in (2+1)D by
dispersion management \cite{abdulla}
and Kerr singularity management \cite{abdul,new,ueda}
as well as in (3+1)D
by  Kerr singularity management \cite{unp1,unp2} in SF medium
has led us to investigate the possibility
of the stabilization of  an optical soliton in (3+1)D by dispersion
management. However, some of these studies were concerned with  the
stabilization of Bose-Einstein condensates \cite{abdul,ueda,unp1} using
the nonlinear Gross-Pitaevskii equation  \cite{rmp} which is very similar
to the
NLS structurally. The extension of the problem of stabilization of
optical soliton  from lower to
higher
dimensions, e.g. from (1+1)D to (3+1)D, is of great interest in the actual
three-dimensional world.  It is also more
challenging due  to the possibility of violent collapse in  SF medium in
higher dimensions \cite{1}.  

We begin the present study with a time-dependent variational solution of
the NLS  with a rapidly    oscillating
dispersion
coefficient in (3+1)D.  The variational method leads to a set of coupled
differential
equations for the soliton width and chirp parameters,  which is solved
numerically by the fourth-order Adams-Bashforth method \cite{koo}. The
variational solution illustrates that a (3+1)D light bullet can be
stabilized over a large propagation distance 
in a Kerr SF medium with a  rapidly  oscillating
dispersion
coefficient for beam power above a critical value. Next we turn to a
complete numerical solution of the   partial
differential NLS   by the Crank-Nicholson method \cite{koo2}. The
complete
numerical solution demonstrates the stabilization of a light bullet 
over large propagation distance by dispersion management in (3+1)D SF
Kerr medium.

We also compare the present stabilization of a light bullet by
dispersion management with the stabilization obtained by applying an
oscillating Kerr nonlinearity as suggested in Ref. \cite{unp2}. 
We find that both schemes leads to comparable stabilization of a light
bullet over a large propagation distance of few thousand units.

Although, the present work is of primary interest in the generation of
robust
optical solitons, it is also of interest from a mathematical point of view
in nonlinear physics in the stabilization of a soliton in three
dimensions. Moreover, this investigation has interesting implication in
the study of Bose-Einstein condensates. The quantum nonlinear equation
governing the evolution of the condensate, known as the Gross-Pitaevskii
equation \cite{rmp}, is identical with the (classical) 
NLS  \cite{1},  for the
evolution of optical soliton, although the interpretation of the different
variables of these two equations  is different. In recent years there has
been experimental  \cite{EX} and theoretical \cite{TH}  studies of
Bose-Einstein condensates in
periodic optical-lattice potentials generated by a standing-wave laser
beam. By an effective-mass description, the Gross-Pitaevskii equation for
the condensate in a periodic optical-lattice potential  can be reduced to
a dispersion-managed NLS, where the
effective
mass could be positive or negative \cite{eff}. There has already been
experimental
application of such dispersion management to Bose-Einstein condensates
\cite{ex}. 
The effective mass can be varied by changing the parameters of the
periodic potential \cite{abdulla}. 
Once a controlled variation of the effective mass would be possible, it
might be possible to stabilize a Bose-Einstein condensate in this fashion.

In Sec. II we present a variational study of the problem and demonstrate 
the possible stabilization of a light bullet in (3+1)D. 
In Sec. III
we present a complete numerical study based on the NLS to study
in detail the stabilization of a dispersion-managed light bullet. We also
compare the present stabilization with that by nonlinearity management as
in Ref. \cite{unp2}. Finally,
in Sec. IV we give the concluding remarks.

\section{Variational Calculation}

For anomalous dispersion, the NLS can be written as \cite{1}
\begin{eqnarray}\label{d1} 
 \biggr[ i\frac{\partial}
{\partial z} +\frac{\gamma(z)   }{2}\nabla_{\bf r}^2    +
|{u({\bf r},z)}|^2
 \biggr]
u({\bf  r},z)=0,
\end{eqnarray}
where in (3+1)D the three dimensional vector ${\bf r}\equiv (x,y,t)$
has space components $x$ and $y$ and time component $t$, and $z$ is the
direction of propagation. The Laplacian operator $\nabla_{\bf r} ^2$ acts
on the
variables $x$, $y$, and $t$. 
The dispersion coefficient $\gamma(z)$ in a Kerr medium is taken
to be rapidly oscillating 
and can have  successive positive  and negative values. 
The normalization condition  is
$ \int d{\bf r} |u({\bf  r},z)|^2 = P$, 
 where $P$ is the power of the
optical beam \cite{st,mal}.

For a 
spherically symmetric 
soliton in (3+1)D,  $u({\bf r},
z)=
U(r,z) $. Then the radial part of the NLS (\ref{d1}) becomes \cite{1}
\begin{eqnarray}\label{d4}
 \biggr[i\frac{\partial
}{\partial z} +\frac{\gamma(z)  }{2}\frac{\partial^2}{\partial
r^2}  +\frac{\gamma(z) }{r}\frac{\partial}{\partial r}           
+  \left|
{U({r},z)}\right|^2 \biggr] U({r},z)=0.
 \end{eqnarray}
For stabilization of the soliton, we shall employ the following
oscillating dispersion coefficient: $\gamma(z)= [1+g_1\sin(\omega z)]$,
where frequency $\omega$ is large.
In this paper we consider  variational and numerical solutions of
Eq.  (\ref{d4}) to illustrate the stabilization of a (3+1)D light bullet.

Recently, it was suggested \cite{unp2} that a rapid variation of the
Kerr
nonlinearity
parameter also stabilizes a soliton. In this study 
we critically compare these two ways of stabilization. For
that purpose we consider 
\begin{eqnarray}\label{d2}
 \biggr[i\frac{\partial
}{\partial z} +\frac{1  }{2}\frac{\partial^2}{\partial
r^2}  +\frac{1 }{r}\frac{\partial}{\partial r}
+n(z)  \left|
{U({r},z)}\right|^2 \biggr] U({r},z)=0,
 \end{eqnarray}
where the nonlinearity parameter will be taken to be rapidly varying for
stabilizing the soliton. In Ref.  \cite{unp2}  $n(z)$ was taken to be
rapidly jumping between positive and negative values and a small
propagation distance $z$ was considered for stabilization. In the present
study a sinusoidal variation of  $n(z)$ is considered over a much larger
propagation distance.  For stabilization we consider $n(z) = [1- g_2 \sin
(\omega z)]$ in this paper, where the second term on the right
is rapidly oscillating
for a large $\omega$ and where $g_2$ is a constant.

First we consider the variational approach with the following trial 
Gaussian wave function 
for the solution of Eq.   (\ref{d4})  
\cite{ueda,abdul,unp2,1}
\begin{equation}\label{twf}
U(r,z)=N(z)   
\exp\left[-\frac{r^2}{2R^2(z)}
+\frac{i}{2}{ b(z)}r^2+i\alpha(z) 
\right],
\end{equation}
where $N(z)$, $R(z)$, $b(z)$, and $\alpha(z)$ are the soliton's amplitude,
width,
chirp, and
phase, respectively. In
Eq. (\ref{twf})  $N(z)={P^{1/2}}/{[\pi^{3/4}R^{3/2}(z)]}$.
 The trial  function   (\ref{twf})  satisfies (a)
the normalization condition  \cite{st,mal}
as well as the boundary conditions
(b) $U(r,z)  \to$ constant  as $r \to 0$ and (c) $|U(r,z)|$ decays
exponentially
as $r \to \infty$ \cite{mal}.
 
The Lagrangian density for
generating Eq.  (\ref{d4})   is \cite{abdul,unp2}
\begin{equation}
{\cal L}(U)=\frac{i}{2}\left(\frac{\partial U}{\partial
z}U^*
- \frac{\partial  U^*}{\partial z} U 
\right)-\frac{\gamma}{2}\left|\frac{\partial
 U}{\partial r} \right|^2   
+\frac{ | U|^4}{2} ,
\end{equation} 
where the functional dependence of the quantities on $r$ and $z$ is
suppressed. 
 The trial  function (\ref{twf}) is
substituted in the Lagrangian density and the 
effective Lagrangian is calculated by
integrating the Lagrangian density: $L_{\mbox{eff}}= \int {\cal L}
(U)
d \bf r.$ With $U$ given by Eq. (\ref{twf}) the effective Lagrangian is
\begin{eqnarray}
L_{\mbox{eff}}&=& \frac{1}{2}\pi^{3/2}N^2(z)R^3(z)
\biggr[ -\frac{3}{2}\dot b(z)
R^2(z)-2
\dot \alpha(z) \nonumber \\   &+& \frac{N^2(z) }{2\sqrt 2} 
-
\frac{3}{2}\frac{\gamma(z)}{R^2(z)}
- \frac{3}{2} R^2(z) b^2(z) \gamma(z)\biggr], 
\end{eqnarray}
where the overhead dot denotes derivative with respect to $z$.
The standard Euler-Lagrange equations for $N(z)$, $R(z)$, $b(z)$, and
$\alpha(z)$ are  \cite{mal,abdul,ueda} given by
\begin{equation}
\frac{d}{d z}\frac{\partial L_{\mbox{eff}}}{\partial\dot \beta}=
\frac{\partial L_{\mbox{eff}}}{\partial \beta}, 
\end{equation} 
where $\beta$ stands for    $N(z)$, $R(z)$, $b(z)$, or $\alpha(z)$.
The Euler-Lagrange equation for $\alpha(z)$ gives the normalization:
\begin{equation}
\pi^{3/2}N^2 R^3 = P.
\end{equation}
The Euler-Lagrange  equations for  $R(z)$ and   $N(z)$ are given by
\begin{equation}\label{x1}
5\dot b +\frac{4\dot
\alpha}{R^2}+\frac{\gamma}{R^4}+5b^2\gamma=\frac{N^2}{\sqrt{2}R^2}
\end{equation}
\begin{equation}\label{x2}
3\dot b +\frac{4\dot
\alpha}{R^2}+\frac{3\gamma}{R^4}+3b^2\gamma=\frac{2N^2}{\sqrt{2}R^2},
\end{equation}
where the $z$ dependence of the variables is suppressed. 
The Euler-Lagrange equation for $b(z)$ is
\begin{eqnarray} \label{el3}
\dot R(z) = R(z) b (z)\gamma(z).\end{eqnarray}
Equations (\ref{x1}) and (\ref{x2}) leads to the following rate of
variation of $b(z)$:
\begin{eqnarray} 
\dot b(z) = \frac{\gamma(z)}{R^4(z)} - b^2(z)\gamma(z) -
\frac{ P}
{2\sqrt{2 \pi^3}}\frac{1}{R^5(z)}. \label{el4} 
\end{eqnarray}
Now defining the new variable $\kappa(z)= b(z)R(z)$,
Eqs. (\ref{el3}) and (\ref{el4}) can be rewritten as 
\begin{eqnarray}\label{el1}
\dot R(z) &=& \gamma(z) \kappa (z) \\
\dot \kappa(z) & = &  \frac{\gamma(z)}{R^3(z)} - \frac{ P}
{2\sqrt{2 \pi^3}}\frac{1}{R^4(z)}. \label{el2}  
\end{eqnarray}  
Equations 
(\ref{el1}) and (\ref{el2}) are the variational  equations of motion 
for  $R(z)$ and  $\kappa(z).$   
Here we consider  the
stability
condition of optical solitons  described variationally by these equations 
 for $\gamma(z)=[1+g_1\sin(\omega z)]$, where the second term is 
rapidly varying  (large $\omega$) with zero mean value. 
However, this set of coupled differential  equations cannot be solved
analytically and we resort to
their numerical solution using the four-step Adams-Bashforth method
\cite{koo}.
For a large $\omega$ and $P$, and  $g_1\ge 2$  stabilization   of
the
soliton  is always possible provided that the initial values of $R(z)$
and $\kappa(z)$ are appropriately chosen. The soliton can also be
stabilized by
employing a periodically oscillating Kerr nonlinearity. However, in that
case the coupled set of differential equations for width $R(z)$ and chirp
$b(z)$ was simpler and 
could  be studied analytically to establish 
stabilization of the system \cite{unp2}.

\begin{figure}
 
\begin{center}
\includegraphics[width=1.00\linewidth]{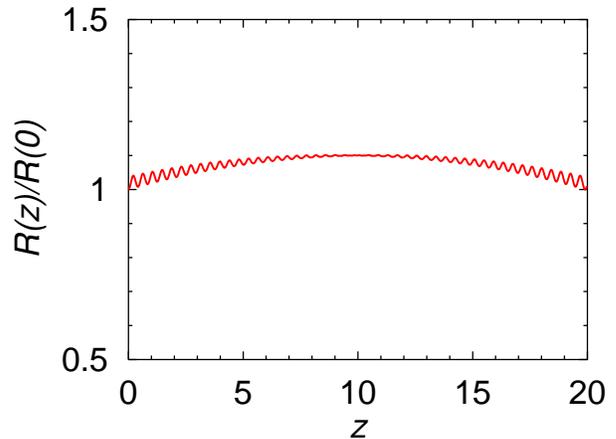}
\end{center}

\caption{The width of the soliton $R(z)/R(0)$ vs. 
propagation distance $z$
from the numerical solution of the coupled set of equations
(\ref{el1}) and
(\ref{el2}), with $\gamma(z)=[1+g_1\sin(\omega z)]$ for $P=500, g_1=4,
\omega=15, R(0)=6$ and $\kappa(0)=0.14$.}
\end{figure}

In Fig. 1 we exhibit the numerical  results for the  width $R(z)$ obtained
from the solution of the variational equations (\ref{el1}) and
(\ref{el2}). In the present calculation we take $\omega =15$, 
$g_1=4$,
$P=500$, and $R(z=0)=6$ If the initial value of $b(z)$ is  chosen
appropriately, stabilization of the system can be obtained. For
these set of parameters, stabilization was obtained for 
$b(z=0)=0.14$, and we show the result in Fig. 1, where we plot
$R(z)/R(0)$ vs. $z$.  The width $R(z)$ is 
found to to remain stable for
a range of values of $z$. Small oscillations in Fig. 1 sustain the soliton
for a large propagation distance $z$.
The
numerical values for $g_1$ and $\omega$ are taken as examples, 
otherwise
they do not have great consequence on the result so long as 
$\omega$ is
large corresponding to rapid oscillation. 
There is no upper limit for  $ P$ and
stabilization seems
possible for an arbitrarily large   $ P$. For larger $P$ the 
stabilization
is more sustained and consequently,
it
is easier to stabilize a soliton.  However, SF 
nonlinearity is essential 
for the stabilization and stabilization is only possible for power $P$ 
above a critical value.

The sinusoidal variation of the dispersion coefficient as considered
above in the  variational study only simplifies the algebra and is by no
means
necessary for stabilization of solitons.  Any  rapid periodic oscillation
of
the dispersion coefficient  was  also found to stabilize the
soliton. 
The same
was found to be true in the stabilization of a soliton by Kerr-singularity
management in (3+1)D \cite{unp1,unp2}.

\section{Numerical Result}

Motivated by the approximate variational study above, now 
we solve Eqs.  (\ref{d4}) and   (\ref{d2})  numerically using  a
split-step iteration method employing the Crank-Nicholson
discretization
scheme  \cite{koo2,11}. 
The $z$-iteration is started with the
known solution of some auxiliary equation
with zero nonlinearity. 
The auxiliary equation with known Gaussian
solution
is obtained by adding
a harmonic oscillator potential $r^2$ to Eqs.  
(\ref{d4}) and   (\ref{d2})  and setting the nonlinear term to zero, e.g.,
\begin{eqnarray}\label{d5}
 \biggr[i\frac{\partial
}{\partial z} +\frac{1  }{2}\frac{\partial^2}{\partial
r^2}  +\frac{1}{r}\frac{\partial}{\partial r}           
+ \frac{d(z)r^2}{4}  \biggr] U({r},z)=0.
 \end{eqnarray}
In Eq. (\ref{d5}) we have introduced a strength parameter $d(z)$ with the
radial trap. Normally, $d(z) = 1$; when the radial trap is switched off
$d(z)$ will be reduced to zero. Asymptotically decaying real boundary
condition for $U(r,z)$ is used throughout this calculation.
 
When the radial trap  is reduced to zero 
the wave function extends to a large value of $r$. Hence to obtain a
converged result a large value of  $r$ is to be considered in the
Crank-Nicholson discretization  
scheme. In the present calculation we employ an $r$-step
of 0.1 with 40 000 grid points spanning an $r$-value up to  4000 and a
$z$-step
of 0.001.  
In the course of $z$-iteration  a 
positive  SF Kerr nonlinearity  with appropriate power $P$
is 
switched on  slowly  and the harmonic trap is also switched off 
slowly.  If the nonlinearity is increased rapidly the system
collapses. The tendency to
collapse or expand to infinity  must be avoided to obtain a stabilized
soliton. 
Although, for the sake of convenience we applied a harmonic trap in the
beginning of our simulation, which is removed later with the increase of
nonlinearity, this restriction is by no means necessary for stabilizing a
soliton.

First we consider the numerical solution of Eq.  
(\ref{d4}) to create the desired soliton.
After switching off the
harmonic trap 
and introducing the final power $P$,
the oscillating dispersion coefficient  $\gamma(z)= [1+g_1\sin(\omega z)]$
is
introduced. 
A stabilization of the final solution could be obtained for a large
$\omega$ and $P$.
If the SF
power after switching off the harmonic trap is large compared to  the
spatiotemporal 
size
of the beam
the system becomes highly attractive in the final stage and it eventually
collapses. If the final power  after switching off the harmonic trap 
is small for its size
the system becomes weakly  attractive in the final stage  and it expands
to infinity. The final power has to have an appropriate
value consistent with the size,   
for final stabilization.

\begin{figure}
 
\begin{center}
\includegraphics[width=1.00\linewidth]{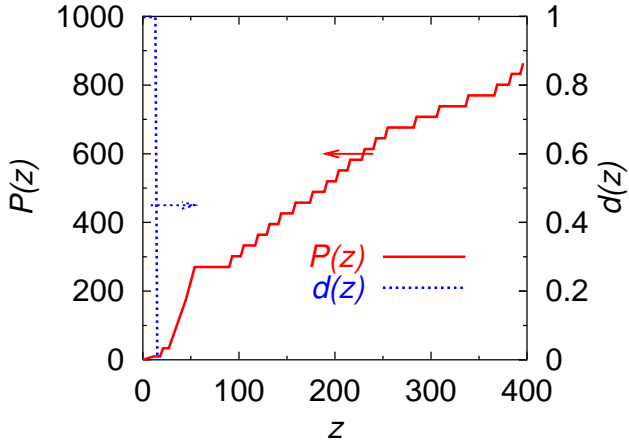}
\end{center}
 
\caption{(Color online) Variation of power $P(z)$ and strength 
$d(z)$ of
radial trap  in the initial stage of stabilization
until the radial trap removed and 
the 
final power $P = 868.4$ attained  at
$z=400$. 
} \end{figure}

\begin{figure}
 
\begin{center}
\includegraphics[width=1.0\linewidth]{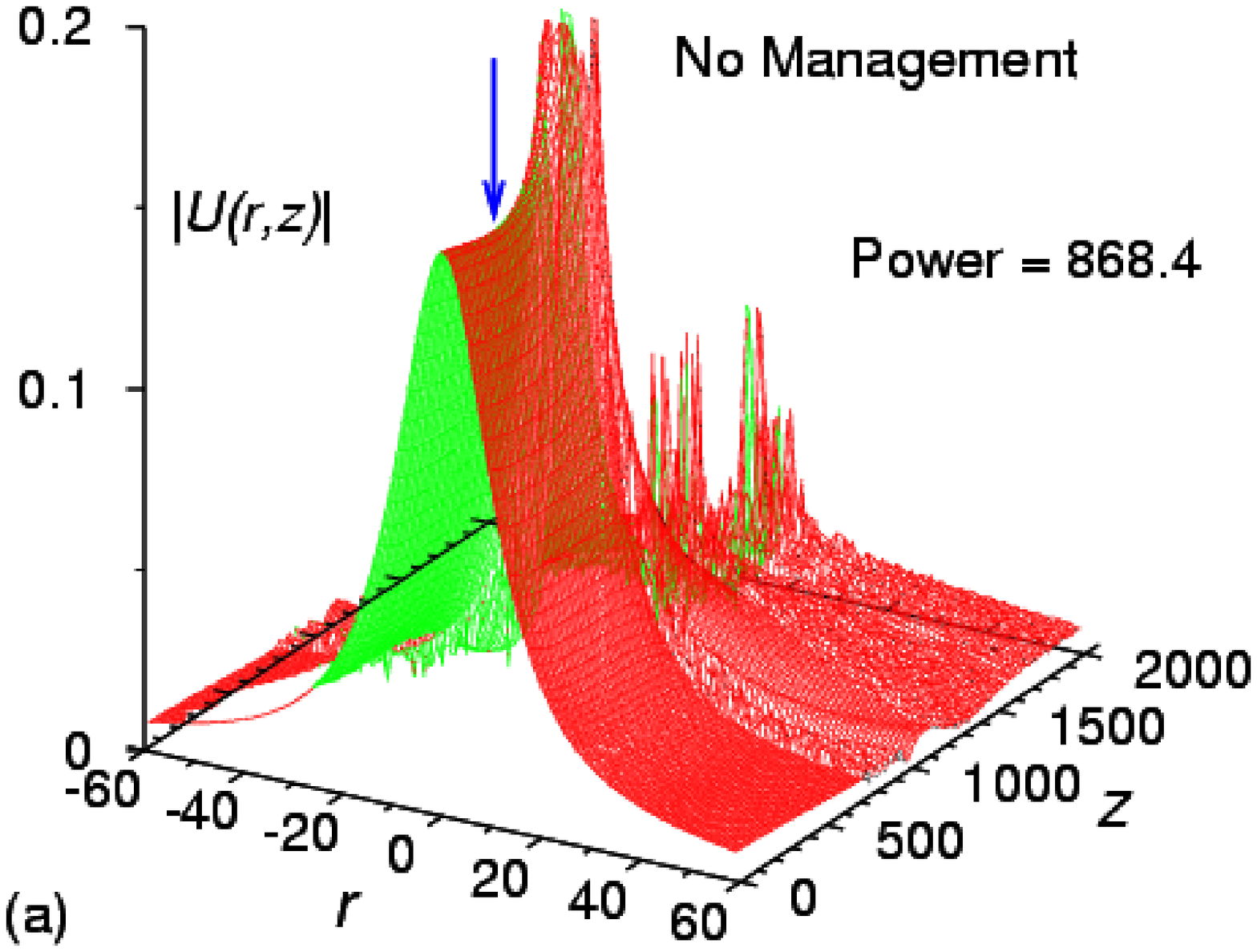}
\includegraphics[width=1.0\linewidth]{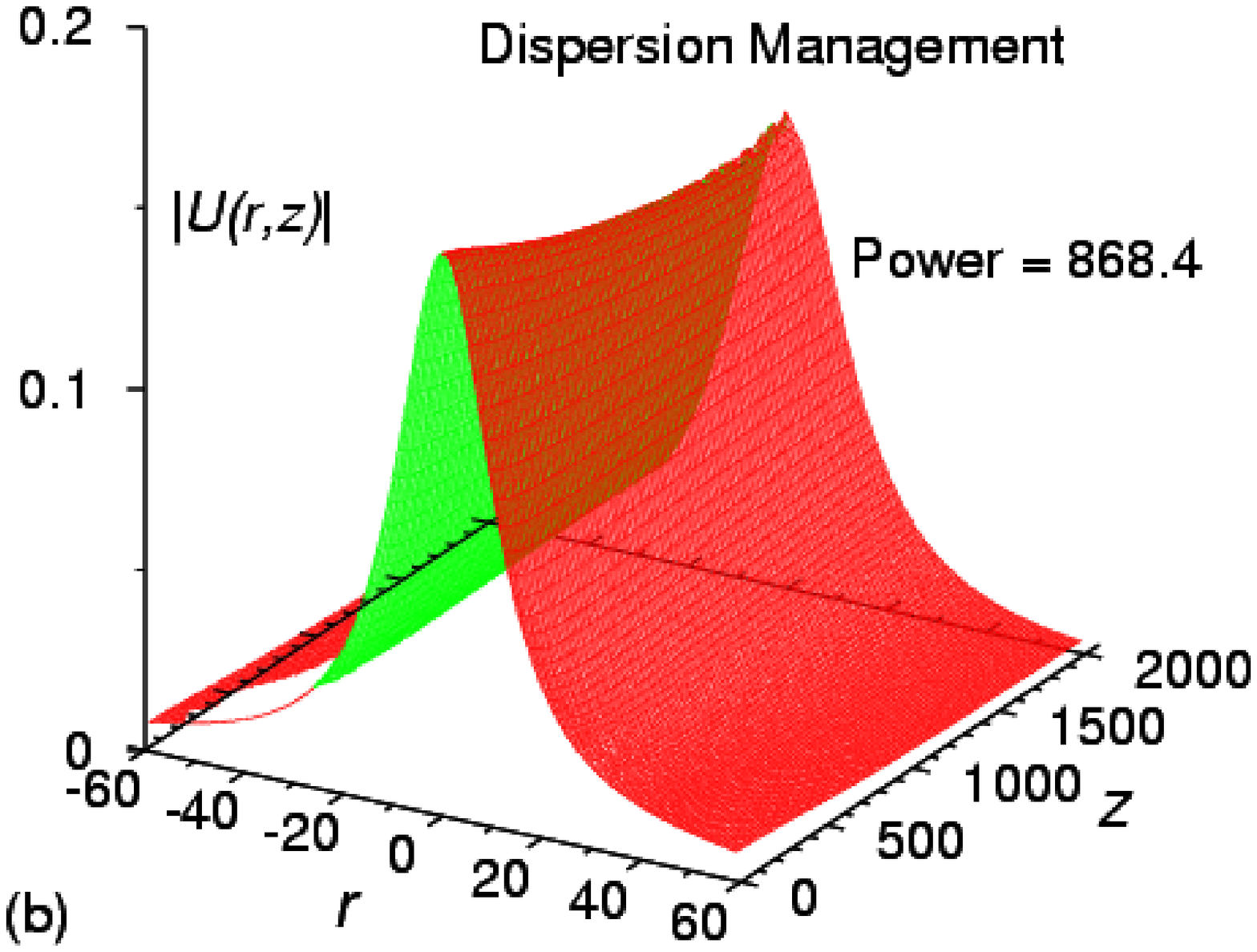}
\includegraphics[width=1.0\linewidth]{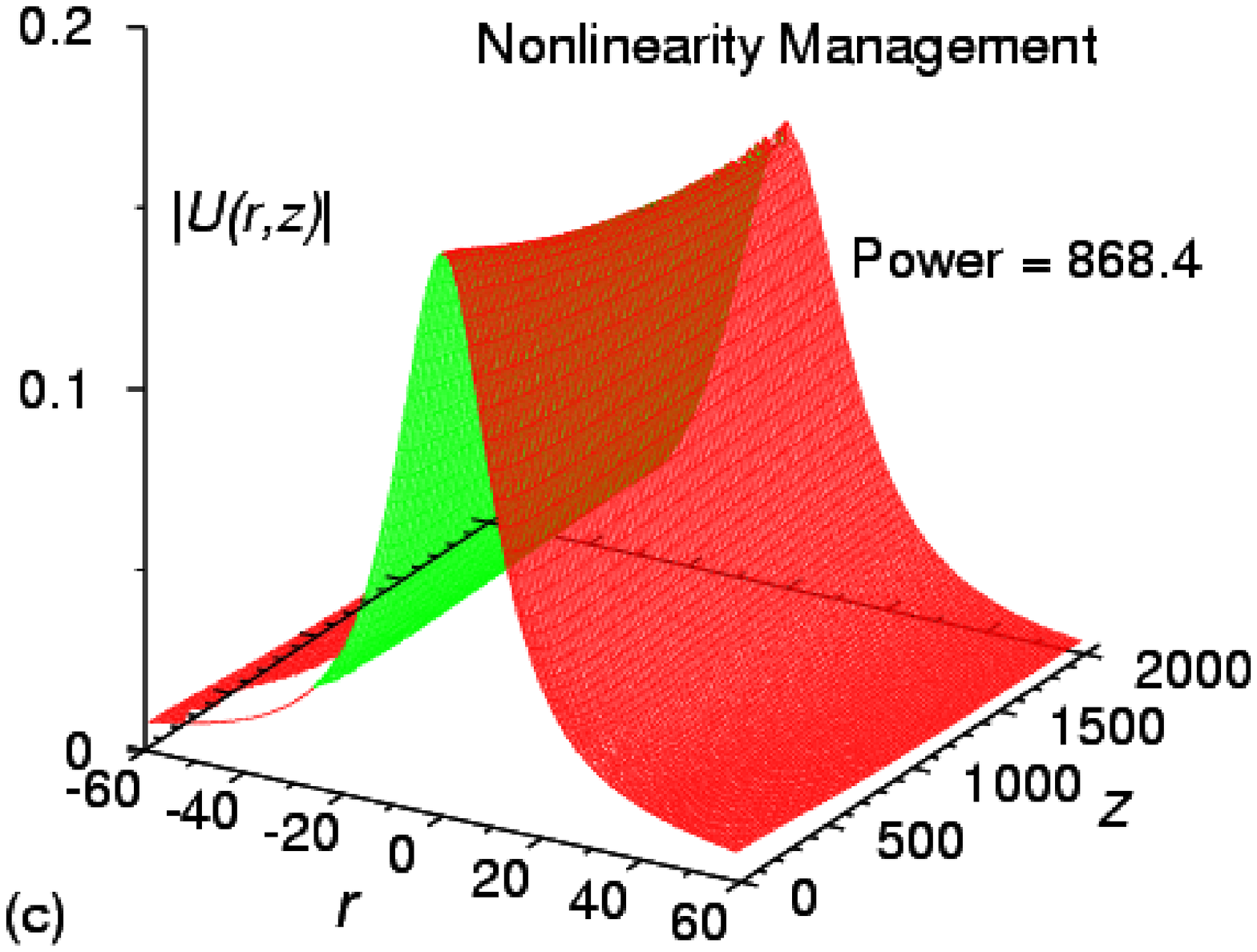}
\end{center}
 
\caption{(Color online) (a) The solitonic wave function $|U(r,z)|$
vs. $r$ and $z$ of a
freely propagating 
light
bullet of power
$P = 868.4$, the arrow at $z=300$ shows the beginning of collapse 
with $|U(r=0,z)|$ starting to diverge; (b) The solitonic wave function
$|U(r,z)|$ of Eq. (\ref{d4}) vs. $r$
and $z$ of  the light bullet of (a) stabilized by dispersion management 
$\gamma(z)=[1+4\sin (5\pi z)]$ applied for all $z>0$; (c)  The solitonic
wave function
$|U(r,z)|$ of Eq. (\ref{d2}) vs. $r$
and $z$ of  the light bullet of (a) stabilized by Kerr nonlinearity  
management
$n(z)=[1-4\sin (5\pi z)]$ applied for all $z>0$. 
}
\end{figure}

Starting from Eq. (\ref{d5})
the soliton can be created with a wide range of variation of power 
$P(z)$ and strength $d(z)$.
In Fig. 2 we show the actual
variation of $P(z)$ and $d(z)$ of the radial
trap in 
Eq. (\ref{d5}). A stabilized soliton can be obtained for different
variations of $P(z)$ and $d(z)$ and the variation in Fig. 2 are by no
means unique and are to be considered as a possible
variation. 
By  
switching off the harmonic trap linearly for $13<z<15$ and 
increasing the nonlinearity  for $0<z<400$, an almost
stationary light
bullet is prepared. One interesting feature of Fig. 2 is that we could not
find a linear variation of $P(z)$
over the whole range to obtain the final soliton. A linear variation
usually led to collapse or expansion to infinity of the final soliton.   
A fine-tuning of the power was needed for obtaining the quasi-stationary
soliton with the kinetic pressure almost balancing the nonlinear
attraction.   This fine-tuning led to the final fractional power
$P=868.4$. 
A  quasi-stationary  soliton, rather than a rapidly  expanding or
collapsing soliton, was ideal for stabilization. Although the final power
$P=868.4$ is used in this study, stabilization can be obtained for any $P$
greater than a critical value. 

In the preparation of the soliton in Fig. 2 no oscillating dispersion
coefficient has been used. The oscillating dispersion coefficient has been
applied later. 
In (2+1)D Saito and Ueda \cite{ueda} first applied a weak
oscillating nonlinearity and then increased its strength in a linear
fashion with $z$ iteration.  The nonlinearity manipulation of  Saito and
Ueda \cite{ueda} is distinct from ours. However, the final result should
be independent of how the  oscillating dispersion or
nonlinearity coefficient is introduced.

The propagation of the free light
bullet
is studied by a direct numerical solution of the NLS (\ref{d4}).
Once it is allowed to propagate in
the $z$ direction the light bullet starts to collapse and 
is destroyed after propagating a
distance of about 300 units of $z$  as shown in Fig. 3
(a) where we plot the profile of the freely propagating light bullet
$|U(r,z)|$ vs. $r$ and
$z$. This means that the nonlinear attraction supersedes the kinetic
pressure  in this soliton and it collapses eventually.

In the present scheme of stabilization the dispersion coefficient  
$\gamma(z)=[1+g_1\sin(\omega z)]$
oscillates rapidly between positive and negative
values. A negative
dispersion coefficient $\gamma(z)$ corresponds to collapse and a
positive
dispersion
coefficient corresponds to expansion. 
In one half of
the oscillating cycle the soliton tends to collapse
and in the other half it tries to expand. 
If, in the first semi-cycle of the application of the 
oscillating 
dispersion coefficient, $\gamma(z)$ is positive, 
the slowly collapsing
soliton of Fig. 3 (a) will tend to expand in this  semi-cycle.
In the next   semi-cycle, $\gamma(z)$ will be  negative and the soliton
will tend to collapse. This happens  for a
positive $g_1>1$.  If     
the expansion  in the first interval is
compensated for by the collapse  in the next interval, a
stabilization of the system is obtained. 
However, if $g_1$ is negative, in the first semi-cycle of
oscillation
the slowly collapsing soliton of Fig. 3 (a)  
will collapse further and a
satisfactory stabilization cannot be obtained. 
For a large $\omega$ the   
system
remains virtually static and the very small oscillations arising from
collapse and expansion remain unperceptible.

\begin{figure}
 
\begin{center}
\includegraphics[width=1.00\linewidth]{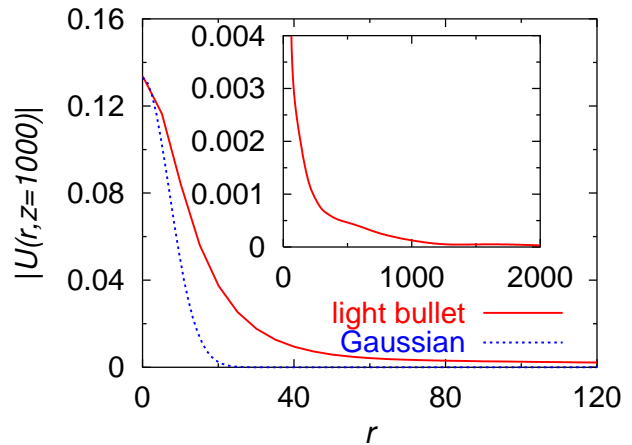}
\end{center}
 
\caption{(Color online) The wave function $|U(r,z=1000)|$ of the
stabilized light bullet of Figs. 3 (b) and (c) vs. $r$. For comparison 
the Gaussian ($\sim \exp(-r^2/100)$) is also shown.} \end{figure}

Next using Eq. (\ref{d4}) we illustrate the possible stabilization of 
the  (3+1)D soliton of power $P=868.4$ of Fig. 3 (a)
by the application of the
oscillating dispersion coefficient $\gamma(z)= [1+g_1 \sin(\omega z)]$
with  $g_1=4$ and $\omega=5\pi$ for $z>0$.
The profile of the dispersion-managed light bullet $|U(r,z)|$ vs. $r$ and
$z$  is plotted in Fig. 3 (b). It is realized that after the application
of dispersion management the soliton remains stable for a large
propagation distance $z$. This
shows clearly the effect of the oscillating dispersion coefficient on
stabilization. 
The dispersion
management of the type considered in Fig. 3  (b) can prolong the life of
the soliton significantly. 
In Fig. 3   (b) the profile  of the solitonic wave
function over the large
interval of $z$ clearly shows the quality of stabilization.
The stabilization seems to be good 
and can 
be continued for longer intervals of  $z$ by considering a soliton of 
larger power.

Finally, using Eq. (\ref{d2}) we consider the stabilization of the above
soliton of power 
$P=868.4$ by the application of an oscillating Kerr nonlinearity 
$n(z)=[1-g_2\sin (\omega z)]$ with $g_2=4$ and $\omega =5\pi$ for $z>0$. 
In this case for $g_2>1$, the Kerr nonlinearity coefficient $n(z)$ is
negative in the first semi-cycle of the oscillating Kerr coefficient. 
This will correspond to an expansion of the soliton, so that the
collapsing tendency of the soliton of Fig. 3 (a) could be stopped and a
stabilization of the soliton could be obtained.  The profile
of the stabilized light bullet by nonlinearity management is plotted
in Fig. 3 (c). It is found that the  oscillating Kerr nonlinearity  can 
stabilize the light bullet for a large propagation distance and enhance
the life of the soliton significantly. The quality of stabilization is
comparable to that obtained by dispersion management. In both schemes the
light bullet can remain stable for few thousands units of $z$. In Figs. 3
we illustrate this stabilization for $z$ up to 2000. The stabilization
obtained in (2+1)D by similar dispersion \cite{abdulla} and nonlinearity
\cite{mal,abdul,new,ueda}
managements was
over a much smaller interval of $z$  ($< 100$).  

The final solitonic wave function $|U(r,z)|$ extends to very  large 
values of $r$. It has a sharp peak for small $r$, reminiscent of strong
binding, and a large extention in $r$, typical of weak binding. We
illustrate these features in Fig. 4 where we plot  $|U(r,z=1000)|$ of a
light bullet
vs. $r$ for dispersion or nonlinearity managements of Figs. 3 (b) and
(c).  The curves for  dispersion and nonlinearity managements    are
identical with each other and a single curve is shown. 
The long tail
of the stabilized light
bullet is shown in inset. 
For comparison a
Gaussian ($\sim \exp(-r^2 /100)$) is also shown.

For both dispersion and nonlinearity managements
we found that the stabilization can only be 
obtained for beams with power larger than a critical value. 
However, we could not obtain a precise value of this critical power 
numerically. In Ref. \cite{unp2} we obtained a variational estimate of
this critical power ($\sim 40$) for nonlinearity management. We could 
not obtain such an estimate for  dispersion  management. 
Numerically, we found that 
it was easier to obtain stabilization of beams with power much larger than
the critical value.

\section{Conclusion}

In conclusion, after a variational and numerical study of the NLS we find
that it is possible to stabilize a spatiotemporal light bullet by
employing a rapidly oscillating dispersion coefficient. We find that the
nature of this stabilization is similar to that obtained by a rapid
variation of the Kerr-nonlinearity parameter. In both cases 
stabilization is
possible in a SF Kerr medium for optical beams with power $P$ larger than
a critical value. However, there is no upper limit of $P$ for
stabilization. A larger value of $P$ is preferred for stabilization. The
stabilization of the light bullet by dispersion and  Kerr nonsingularity
management 
\cite{unp2} should find experimental application in optics and
Bose-Einstein condensation.  Such a dispersion- and nonsingularity-managed
optical soliton can
propagate large distances with a minimum of distortion and is to be
preferred over normal solitons in optical communication.

\acknowledgments

The work was supported in part by the CNPq 
of Brazil.

\end{document}